\documentclass[12pt]{article}

\usepackage{newtxtext,newtxmath}

\usepackage{graphicx}

\usepackage[letterpaper,margin=1in]{geometry}

\renewenvironment{abstract}
	{\quotation}
	{\endquotation}

\date{}

\makeatletter
\renewcommand{\fnum@figure}{\textbf{Figure \thefigure}}
\renewcommand{\fnum@table}{\textbf{Table \thetable}}
\makeatother

\usepackage{scicite}

\usepackage{url}

\newcommand{\ket}[1]{\ensuremath{|{#1}\rangle}}

\usepackage{booktabs}

\def\scititle{
Mechanical Resonator-based Quantum Computing
}
\title{\bfseries \boldmath \scititle}

\author{
	Yu~Yang$^{1,2,\dagger}$,
	Igor~Kladarić$^{1,2,\dagger,*}$,
	Martynas Skrabulis$^{1,2}$,
    Michael Eichenberger$^{1,2}$,\and
    Stefano~Marti$^{1,2}$,
    Simon Storz$^{1,2}$,
    Jonathan Esche$^{1,2}$
    Raquel García Bellés$^{1,2}$,\and
    Max-Emanuel Kern$^{1,2}$,
    Andraz Omahen$^{1,2}$,
    Arianne Brooks$^{1,2}$,\and
    Marius Bild$^{1,2}$,
    Matteo Fadel$^{1,2}$,
    Yiwen~Chu$^{1,2,*}$
\\
	\small$^{1}$Department of Physics, ETH Zürich, 8093 Zürich, Switzerland\and 
    \small$^{2}$Quantum Center, ETH Zürich, 8093 Zürich, Switzerland\and
	\small$^\ast$Corresponding authors -  Email: ikladaric@phys.ethz.ch; yiwen.chu@phys.ethz.ch\and
	\small$^\dagger$These authors contributed equally to this work.
}

\begin{document}

\maketitle
\begin{abstract} \bfseries \boldmath
Hybrid quantum systems combine the unique advantages of different physical platforms with the goal of realizing more powerful and practical quantum information processing devices.
Mechanical systems, such as bulk acoustic wave resonators, feature a large number of highly coherent harmonic modes in a compact footprint, which complements the strong nonlinearities and fast operation times of superconducting quantum circuits. Here, we demonstrate an architecture for mechanical resonator-based quantum computing, in which a superconducting qubit is used to perform quantum gates on a collection of mechanical modes. We show the implementation of a universal gate set, composed of single-qubit gates and controlled arbitrary-phase gates, and showcase their use in the quantum Fourier transform and quantum period finding algorithms. These results pave the way toward using mechanical systems to build crucial components for future quantum technologies, such as quantum random-access memories.

\end{abstract}
\maketitle 

\section*{Introduction}
Quantum information processing (QIP) promises significant advantages over classical computing for a multitude of problems, such as quantum simulations\cite{feynman2018simulating, lloyd1996universal}, unstructured search\cite{grover1996fast}, machine learning\cite{biamonte2017quantum}, and prime factorization\cite{shor1999polynomial, lucero2012computing}. Among the various platforms proposed for the realization of QIP, superconducting circuits have recently shown crucial developments in system scaling and quantum error correction\cite{kim2023evidence, google2025quantum}. However, they still face key challenges, such as limited system connectivity and the lack of a clear separation between processing and memory units, an architectural feature routinely exploited in classical computing\cite{kjaergaard2020superconducting}. A promising approach to overcome these limitations introduces quantum memories -- subsystems containing a large number of modes with long coherence times that can be used for temporary storage of quantum information in conjunction with a quantum central processing unit. 

Electromagnetic realizations of quantum memories couple superconducting qubits to arrays of harmonic oscillators, such as coplanar waveguide resonators\cite{naik2017random, mariantoni2011implementing} and 3D cavities\cite{milul2023superconducting, li2025cascaded}. The long coherence times and high-fidelity storage operations of these systems satisfy the preliminary requirements for building quantum memories. However, their electromagnetic nature results in bulkiness and diminished storage capacity due to access to a limited number of modes. An alternative approach involves a hybrid quantum architecture\cite{pechal2018superconducting, chu2020perspective} in which superconducting qubits are instead coupled to mechanical systems, such as phononic crystal defects\cite{wollack2022quantum}, surface-acoustic wave resonators\cite{satzinger2018quantum} and high-overtone bulk-acoustic wave resonators (HBARs)\cite{chu2017quantum}. HBARs in particular present unique advantages for quantum memory applications, such as long intrinsic coherence times\cite{luo2025lifetime}, compactness, dense multi-mode spectra, and the potential for coherent coupling to other long-lived quantum systems, such as spin qubits\cite{ovartchaiyapong2014dynamic, barfuss2015strong}. Furthermore, 
the ability to implement both single- and two-qubit gates enables the usage of coupled qubit-resonator systems as viable universal quantum computing platforms on their own\cite{pechal2018superconducting, mariantoni2011implementing}.

\begin{figure}
\centering
\includegraphics[width=16cm]{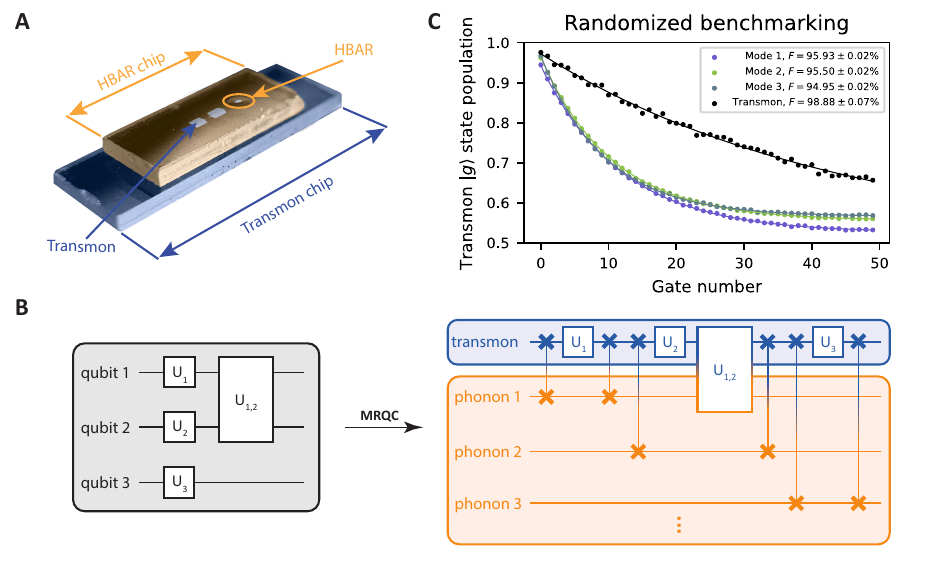}
\caption{
\textbf{Operating principle of MRQC and randomized benchmarking.} 
\textbf{A}, Image of the $\hbar$BAR device. The transmon qubit is fabricated on the bottom chip (false-colored blue), while the mechanical modes reside within the HBAR chip on top (false-colored orange).
\textbf{B}, General principle of the MRQC architecture. A generic conventional quantum circuit containing single- and two-qubit gates is depicted on the left, while the equivalent circuit implemented in MRQC is shown on the right. 
\textbf{C}, Randomized-benchmarking results for single-phonon gates for three different modes (colors) compared to single-transmon gates (black).
}
\label{fig1}
\end{figure}

In this work, we realize a mechanical resonator-based quantum computing (MRQC) platform  using an $\hbar$BAR device -- a circuit quantum acoustodynamics (cQAD) system composed of a superconducting transmon qubit coupled to an HBAR, depicted in Fig.~\ref{fig1}A. This architecture leverages the fast, high-fidelity single-qubit control and readout of the superconducting transmon qubit, together with the multi-mode structure and long coherence time of the phonon modes in the HBAR. We demonstrate the implementation of a set of both single- and two-qubit gates on states stored in the mechanical modes, sufficient for universal quantum computing. In particular, we introduce a scheme for implementing controlled arbitrary-phase ($C_{\phi}$) gates via off-resonant qubit–resonator interactions and employ it to realize commonly used quantum algorithms such as the quantum Fourier transform (QFT)\cite{nielsen2010quantum}.

In our system, the interactions between the mechanical modes of the HBAR and the transmon qubit are governed by the Jaynes-Cummings (JC) Hamiltonian

\begin{equation}
H/\hbar = \frac{\omega_q}{2} \sigma_z + \sum_i{\omega_{i} {a_i}^\dagger a_i} + \sum_i{g\left( \sigma_+ a_i + \sigma_- {a_i}^\dagger\right)},     
\end{equation}
where $\sigma_-$ and $a_i$ are annihilation operators of the qubit and the mechanical (phonon) mode $i$, $\omega_q$ and $\omega_{i}$ are their frequencies, and $g/2\pi = 296\,\text{kHz}$ is the coupling strength between the qubit and each phonon mode. An additional, far-detuned microwave tone is applied to induce an AC Stark-shift on the qubit, allowing us to control its frequency\cite{chu2017quantum}. In MRQC, the transmon qubit functions as the central processing unit (CPU), while the multitude of modes of a single HBAR serve as a random-access memory (RAM), analogous to a classical computer. An example illustrating this idea and its equivalency to a conventional quantum circuit is shown in Fig.~\ref{fig1}B. The general protocol is as follows: phonon modes 1, 2, and 3 store the quantum states of qubits 1, 2, and 3, respectively, in the subspace of their $\ket{0}$ and $\ket{1}$ Fock states. To apply operations on the states stored in the phonon modes, an iSWAP gate is applied between a mode and the transmon via a resonant JC interaction\cite{chu2017quantum}. As detailed below, a single- or a two-qubit gate is then applied by driving the transmon, or by inducing off-resonant JC interactions with another phonon mode, respectively. Finally, the resulting state in the transmon is swapped back into the original phonon mode.

\section*{Single- and two-qubit gates}

We now demonstrate a set of single- and two-qubit gates with our system that are required for universal quantum computing. Single-phonon gates consist of a swap operation, a single-transmon gate, and a final swap operation. To characterize their fidelity, we apply a randomized benchmarking (RB) protocol \cite{knill2008randomized} to a set of gates belonging to the Clifford gate set for three different phonon modes, labeled with indices 1 to 3, and measure fidelities of $95.93\%$, $95.50\%$ and $94.95\%$ respectively. By comparing these results with the RB data for pure single-transmon gates in Fig.~\ref{fig1}C, we extract that a single swap operation induces an infidelity of 1.71\% on average across all three modes which is in reasonable agreement with the predicted value given the coherence properties of the device (see Supplementary Information B).

\begin{figure}
\centering
\includegraphics[width=16cm]{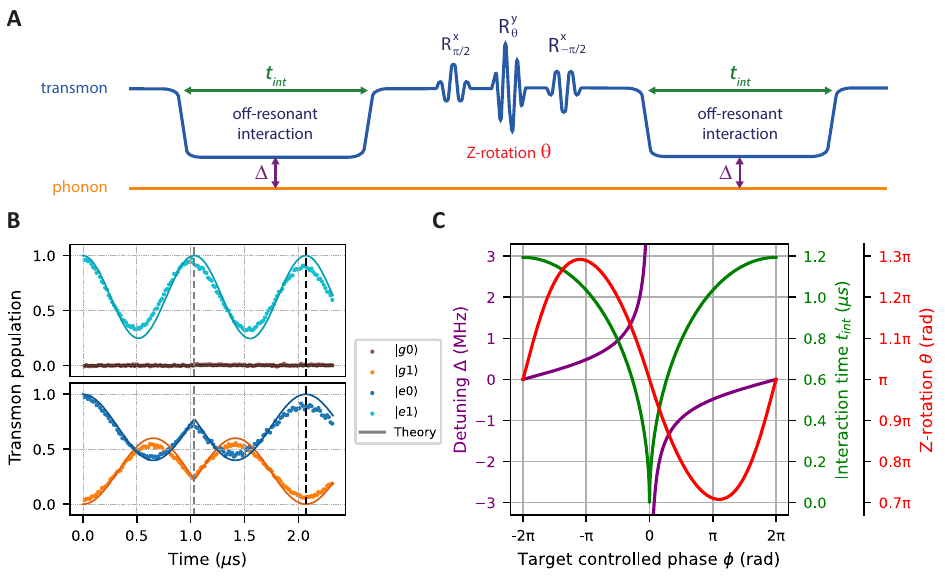}
\caption{ 
\textbf{Arbitrary controlled-phase gate protocol.} 
\textbf{A}, Sequence for implementing a $C_{\phi}$ operation composed of two off-resonant transmon-phonon interactions and a Z-rotation operation in the middle. The Z-rotation is decomposed into a series of X- and Y-rotations.
\textbf{B}, Measured and theoretical evolution of the transmon \ket{e} state populations during the $C_{\phi}$ gate for four different initial states. Gray and black vertical dashed lines show the point at which the Z-rotation is applied and the point at which the $C_{\phi}$ gate is completed, respectively.
\textbf{C}, Dependence of the detuning $\Delta$ (purple), interaction time $t_{int}$ (green) and Z-rotation $\theta$ (red) on the target controlled phase $\phi$ (see Supplementary Information). For later analysis of measured gate fidelities, note that a larger $\phi$ require a smaller $\Delta$ and a longer $t_{int}$. 
}
\label{fig2}
\end{figure}

To enable two-qubit gates, we developed a protocol for applying controlled arbitrary-phase ($C_{\phi}$) gates between the transmon and a phonon mode. The protocol is inspired by the phase accumulation that is induced during off-resonant transmon–phonon interactions in the dispersive regime \cite{vonLupke22}. The experimental sequence, shown in Fig.~\ref{fig2}A, consists of an off-resonant interaction with transmon–phonon detuning $\Delta$ for a duration $t_{int}$, followed by a single-transmon Z rotation with phase $\theta$, and concludes with the same off-resonant interaction as in the first step.

 A key requirement for properly implementing a $C_{\phi}$ gate is that the populations in the computational basis states return to their initial values by the end of the operation, ensuring only phase accumulation for all four initial states $\ket{g0}$, $\ket{e1}$, $\ket{g1}$ and $\ket{e0}$, as can be seen in Fig.~\ref{fig2}B. Therefore, one way to understand how our protocol works is by examining the evolution of the transmon population during the application of the $C_{\phi}$ sequence for a given initial state, as illustrated in Fig.~\ref{fig2}B for the case of $\phi = \pi$. In the trivial case where both subsystems are initialized in the ground state $\ket{g0}$, we observe no oscillations in the transmon state populations, because none of the involved operations introduce additional excitations to the system. In the second case, where the system is initialized in the doubly excited state $\ket{e1}$, the system will oscillate between the $\ket{e1}$ and $\ket{g2}$ states with frequency $\omega_2 = \sqrt{\Delta^2 + 8g^2}$. We choose the off-resonant interaction time $t_{int} = 2\pi/\omega_2$ at which the transmon returns to the $\ket{e}$ state. This way, regardless of the chosen phase $\theta$ of the applied Z-rotation, the $\ket{e1}$ population again returns to its initial value after another off-resonant interaction with duration $t_{int}$.

In the remaining two cases, where the system is initialized in the $\ket{g1}$ or $\ket{e0}$ state, the system oscillates between the two states with a slower frequency of $\omega_1 = \sqrt{\Delta^2 + 4g^2}$, and therefore the state of the system is neither $\ket{g1}$ nor $\ket{e0}$ after interaction time $t_{int}$. This means that the choice of the Z-rotation phase $\theta$ has an effect on the final state of the system after the second off-resonant interaction. As a result, a unique triplet of parameters $\Delta$, $t_{int}$ and $\theta$ exists for a target controlled-phase $\phi$, for which an operation

\begin{equation}
    \begin{pmatrix}
        1&0&0&0\\
        0&e^{i\phi_1}&0&0\\
        0&0&e^{i\phi_2}&0\\
        0&0&0&e^{i(\phi_1+\phi_2+\phi)}\\
    \end{pmatrix}
\end{equation}

\noindent is implemented, as shown in Fig.~\ref{fig2}C. Analytical expressions for the parameters are given in the Supplementary Information E. This gate is equivalent to a $C_{\phi}$ operation, up to single-qubit rotation phases $\phi_1$ and $\phi_2$, which can be taken into account for subsequent single-qubit operations. For a set of gate parameters chosen such that a $C_{\pi}$ gate is implemented, we experimentally measure the evolution of the transmon $\ket{e}$ state population and confirm that it returns to its initial value for every initial basis state, as shown in the scatter plots of Fig.~\ref{fig2}B.  

\section*{Process tomography of two-qubit gates}
\begin{figure}
\centering
\includegraphics[width=16cm]{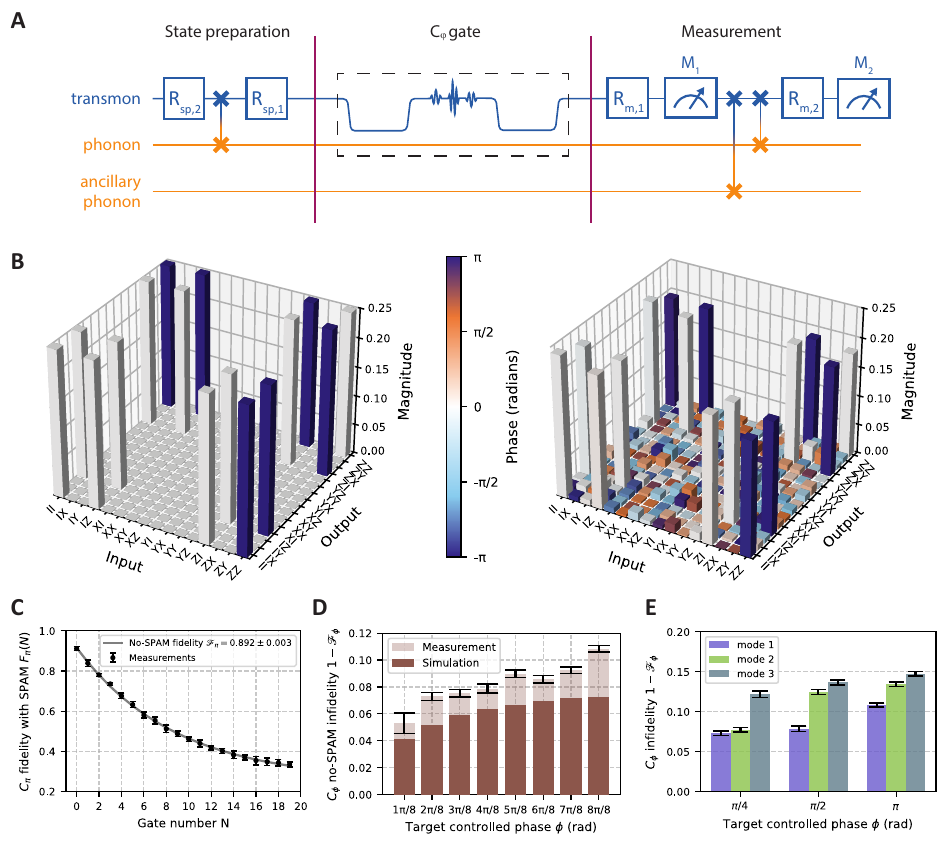}
\caption{
\textbf{Arbitrary controlled-phase gate process tomography.} 
\textbf{A}, Process tomography sequence of an arbitrary $C_{\phi}$ gate. The choice of initial states and measurement axes are determined by $R_{sp}$ and $R_m$ pulses, respectively.
\textbf{B}, Ideal (left) and reconstructed (right) process matrices $\chi$ of a $C_{\pi}$ gate. The implemented $C_{\pi}$ gate yields a fidelity $F_{\pi} = 85.7\%$ including SPAM errors.
\textbf{C}, Measured fidelities of multiple repetitions of $C_{\pi}$ gates with SPAM errors. The exponential fit to the data allows us to extract the no-SPAM fidelity of $\mathscr{F}_{\pi}=89.2\%$. Error bars are determined by repeating the experiment ten times and calculating the standard deviation. 
\textbf{D}, Measured (transparent) and simulated (opaque) no-SPAM infidelities $1-\mathscr{F}_{\phi}$ of eight $C_{\phi}$ gates with $\phi = k \pi/8$.
\textbf{E}, Measured $C_{\pi/4}$, $C_{\pi/2}$ and $C_{\pi}$ gate infidelities $1-\mathscr{F}_{\phi}$ for three different phonon modes (see Supplementary Information A for details on the modes' properties).
}
\label{fig3}
\end{figure}

To show that the protocol indeed implements $C_{\phi}$ gates and to characterize the gate fidelities, we perform quantum process tomography of the transmon–phonon controlled-phase operation, as shown in Fig.~\ref{fig3}A. For this, we prepare each subsystem—the transmon and the phonon mode—in the states $\ket{0}, \ket{1}, \ket{+}$ and $\ket{i}$. Taking the tensor product of these states yields a complete set of 16 product input states. For each input state, we apply the operation we wish to characterize and finally measure the states of the transmon and the phonon mode along all 9 pairs of axes $x$, $y$ and $z$. In total, the choice of input states and measurement axes yields 144 different measurements from which we reconstruct the output density matrix of the system using single-shot readout results corrected for readout infidelity (see Supplementary Information D). Fig.~\ref{fig3}B shows an example of the measured $\chi$ matrix for $\phi = \pi$ on phonon mode 1, together with the corresponding ideal operation. From this, we extract a fidelity of $F_{\pi} = 85.7\%$, which includes contributions from state-preparation-and-measurement (SPAM) errors. The dominant measurement errors arise from phonon decay during the qubit readout time and the additional SWAP operation required to read out the phonon state.

To determine the gate fidelity without SPAM errors, we employ the process tomography protocol for an operation consisting of N repetitions of the $C_{\pi}$ gate, with N ranging from 0 to 19. Fitting an exponential function $F_{\pi}(N) = A\mathscr{F}_{\pi}^N+B$ to the obtained results allows us to extract the single $C_{\pi}$ gate fidelity of $\mathscr{F}_{\pi}=89.2\%$, as seen in Fig.~\ref{fig3}C. 

We repeat the same procedure for controlled-phase gates with target phases that are integer multiples of $\pi/8$ and compare the results with simulations of the same protocol, as shown in Fig.~\ref{fig3}D. We observe that smaller target phases exhibit lower gate infidelities, which is expected since smaller $\phi$ values require shorter interaction times $t_{\mathrm{int}}$, leading to less decay and decoherence. The discrepancy between simulation and experiment may arise from unwanted transmon coupling to additional phonon modes with higher transverse mode numbers and other effects that are not taken into account in the simulations, such as variability of transmon properties with frequency and dephasing caused by noise on the AC Stark-shift drive.

In Fig.~\ref{fig3}E, we present and compare the infidelities of the $C_{\pi/4}$, $C_{\pi/2}$, and $C_{\pi}$ gates for three different phonon modes, as these gates are used in the algorithms described below. All measured gate fidelities exceed $85\%$, with modes with higher coherence properties and lower required AC Stark-shifts exhibiting higher fidelities, as expected.

Our $C_{\phi}$ gate protocol offers several advantages: (1) it enables arbitrary controlled phases, up to single-qubit rotation corrections; (2) it is fast — even the $C_{\pi}$ gate, which requires the longest interaction time, has a total duration of only about $2.5$ times the SWAP gate duration, which is approximately 10 times faster than the decoherence time of the transmon; and (3) it is also applicable to other systems consisting of a two-level system coupled to a harmonic oscillator via a JC interaction, without requiring an auxiliary non-computational states of the transmon, unlike other implemented protocols for arbitrary controlled-phase gates\cite{barends2015digital, cscarato2025, collodo2020implementation}.

\section*{Quantum Fourier transform}
We take advantage of the introduced $C_{\phi}$ gates to implement a quantum Fourier transform algorithm. The QFT is a crucial part of many QIP algorithms, such as Shor's prime factorization, quantum phase estimation, and other algorithms that require period finding in structured data\cite{nielsen2010quantum}. It is particularly interesting for demonstrating the capabilities of MRQC because it relies on the ability to apply controlled arbitrary-phase gates, full connectivity of the circuit, as well as the fact that, although a QFT circuit on $n$ qubits contains a total of $\Theta(n^2)$ gates, each individual qubit is only operated on $O(n)$ times and otherwise remains idle. The last point highlights the advantage of the MRQC platform because, during the idling times, the states are stored in the phonon modes with significantly longer coherence times compared to the transmon.

\begin{figure}
\centering
\includegraphics[width=16cm]{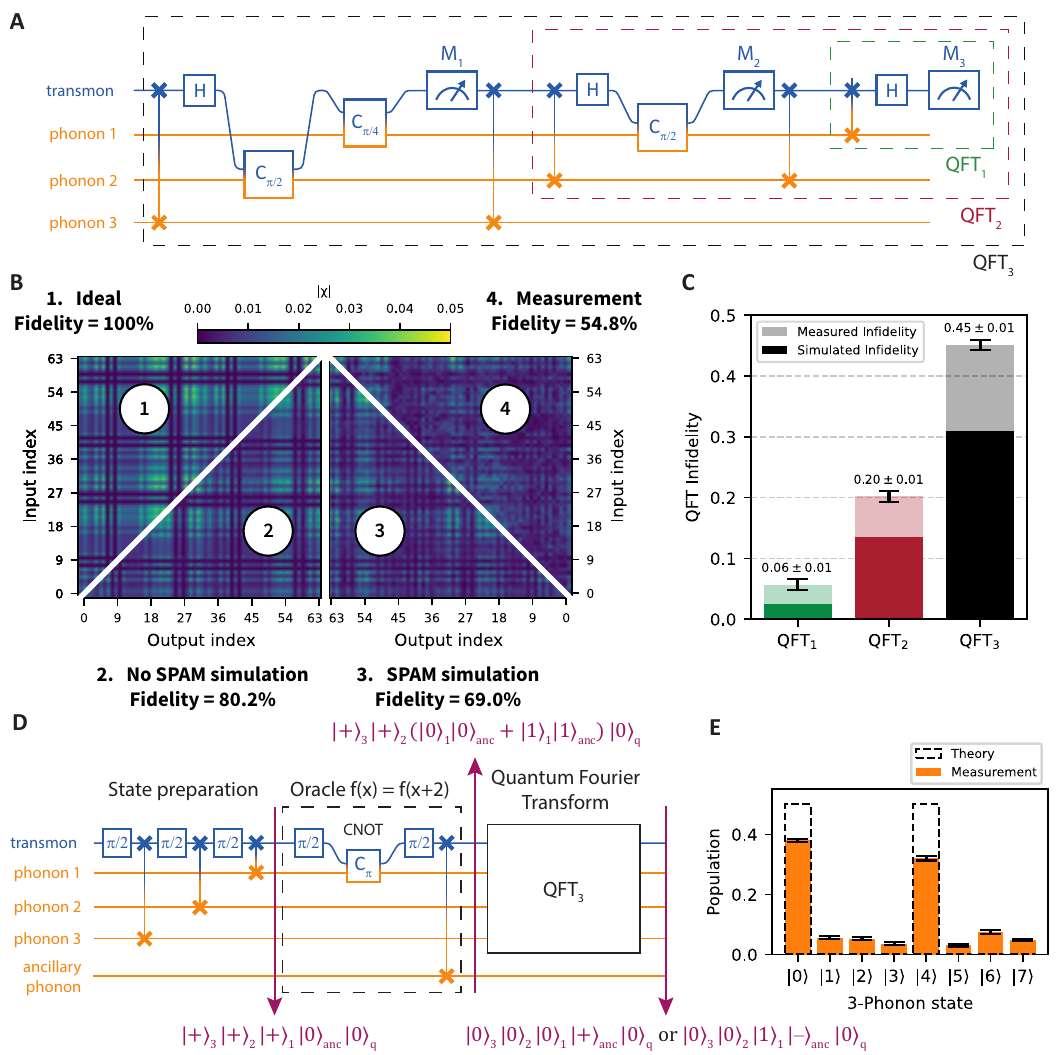}
\caption{ 
\textbf{Quantum Fourier transform and quantum period finding measurements.} 
\textbf{A}, QFT sequence for one (green), two (red) and three (black) phonon modes.
\textbf{B}, Reconstructed $\chi$ matrix amplitudes of a three phonon QFT sequence for an ideal operation (1), simulation without SPAM (2), simulation with SPAM (3), and experimentally measured (4). Index labels on x- and y-axes correspond to combinations of 3-element tensor products of Pauli operators (see Supplementary Information D). 
\textbf{C}, Measured (transparent) and simulated (opaque) infidelity of the QFT operations with SPAM.
\textbf{D}, Sequence for a quantum period finding algorithm on a function with period $r=2$. Ideal intermediate states during the algorithm are shown in pink.
\textbf{E}, Measured population results of the QPF algorithm shown in D. We observe clear population peaks in states $\ket{0}$ and $\ket{4}$, corresponding to $r=2$. 
}
\label{fig4}
\end{figure}

We use the MRQC circuit depicted in Fig.~\ref{fig4}A to perform the QFT protocol on the space spanned by the basis states $\{\ket{x=0}, ...,\ket{7} \}$. First, the input state $\ket{\psi}=\sum_{x=0}^7c_x\ket{x}$ is prepared in three phonon modes using binary encoding, while the transmon is initialized in the ground state. Then we swap the state of phonon mode 3, which encodes the most significant digit, into the qubit and apply a Hadamard gate followed by $C_{\pi/2}$ and $C_{\pi/4}$ gates with phonon modes 2 and 1, respectively. After this point in the sequence, one would in principle swap the state of the transmon back to phonon mode 3, which will remain idle for the rest of the protocol. However, since we are not using the QFT output state for subsequent calculations, we simply measure the transmon state instead to avoid additional decoherence due to mode 3. Here, the measurement M involves a qubit pulse that determines the axis in which we measure the state, as well as dispersive state readout. Additionally, we wait for $3 \text{ $\mu$s}$ after each measurement for the readout resonator to depopulate before proceeding with the rest of the sequence (see Supplementary information D). We then reset the transmon by swapping its projected state into phonon mode 3, which is no longer used. We continue with the QFT protocol on the remaining phonon modes 2 and 1 as shown in the figure. 

We repeat this procedure for 64 different initial states and 27 different choices of measurement axes to perform process tomography on the QFT protocol (see Supplementary Information D). The amplitudes of the extracted process matrix $\chi_{QFT, 3}$ and its comparison with the ideal QFT operation, as well as simulations of the same procedure with and without SPAM errors are shown in Fig.~\ref{fig4}B. We observe qualitative agreement between the measured and expected processes, with measured fidelity with SPAM of $F_{QFT,3} = 54.8\%$. Simulations of the protocol show that approximately $20\%$ of the infidelity is attributed to system decoherence and $11\%$ to SPAM errors. The remaining discrepancy between measurements and simulations can be explained by imperfect system calibration, effects of nearby higher-order transverse modes, and simplification of the readout process in simulated results (see Supplementary Information D). For comparison, we additionally implement the QFT protocol for $n=2$ and $n=1$ phonon modes, shown in red and green in Fig.~\ref{fig4}A. Results of these measurements also show discrepancy with the simulations, however to a lesser extent due to shorter sequence durations resulting in smaller accumulation of previously mentioned effects unaccounted in the simulations. The process infidelities for all three system dimensions are shown in Fig.~\ref{fig4}C. 

Here we make two important comments about this implementation of the QFT. First, it reverses the bit significance order. For example, the state $\ket{4}$ stored in the phonon modes before the QFT sequence is represented as $\ket{1}_3\ket{0}_2\ket{0}_1$, while at the end of the algorithm, it is encoded as $\ket{0}_3\ket{0}_2\ket{1}_1$ (see Supplementary Information E). Second, making mid-circuit measurements means that we are not performing tomography on a particular output state that exists at the end of the protocol. Practically speaking, both points are not an issue for algorithms where the QFT is the last step, and we are therefore interested only in the classical results \cite{griffiths1996semiclassical, chiaverini2005implementation}.

\section*{Quantum period finding}
After demonstrating the QFT, we use it to implement an example of the quantum period-finding (QPF) algorithm, which acts as an essential subroutine in Shor's algorithm\cite{shor1999polynomial} and further showcases the potential of the MRQC architecture. The QPF circuit diagram using three phonon modes is shown in Fig.~\ref{fig4}D. We first prepare the phonon modes in the superposition state $\ket{+}_3\ket{+}_2\ket{+}_1$, and then apply the oracle of a function
\[
f(x) =
\begin{cases}
0, & \text{if $x$ is even}, \\
1, & \text{if $x$ is odd},
\end{cases}
\]
followed by application of the QFT protocol introduced earlier.  The function $f(x)$ has a period of $2$ and its oracle is implemented by a CNOT gate between the qubit and phonon mode~1. The CNOT gate is decomposed into single-qubit rotations and a $C_{\pi}$ gate. After the CNOT operation, the qubit state is swapped with an ancillary phonon mode. This additional SWAP operation ensures that the state stored in the phonon modes remains entangled with the state in the ancillary mode, while the transmon is subsequently left in the ground state, enabling the application of the QFT protocol. Effectively, the SWAP operation traces out the state of the transmon before proceeding with the protocol. Finally, we apply the QFT circuit and measure the resulting populations in the basis states, which are shown in Fig.~\ref{fig4}E. We observe clear peaks in the populations for the output states $\ket{0}$ and $\ket{4}$, indicating a period of $r = 2$, which we confirm quantitatively using classical post-processing (see Supplementary Information G).

\section*{Conclusion and Outlook}
We have demonstrated a new architecture for quantum computing that takes advantage of the unique properties of mechanical resonators. Our implementation of a universal set of single- and two-qubit gates can be generalized to any platform involving resonator modes coupled to a qubit through a JC interaction, including other cQAD systems. While we have demonstrated proof-of-principle MRQC operations on three phonon modes in this work, the dense multi-mode spectral structure of the mechanical resonators provides many more modes that can be used as long-living quantum memories, with an all-to-all connectivity via the frequency-tunable transmon qubit. 

Currently, the size of the implementable quantum processor is limited by the number of modes with which the transmon can resonantly interact, while the length of quantum circuits is mainly limited by the electromechanical coupling strength compared to the coherence times and transmon readout duration. While we are making steady improvements to the system coherence, another useful development for future demonstrations would be to move to a coplanar circuit platform\cite{kervinen2020sideband}. This would allow for fast flux-tuning of the transmon and access to a significantly higher number of phonon modes for increased storage capacity. Furthermore, the addition of on-chip Purcell filters would enable faster transmon readout, resulting in shorter phonon idling times and improved operation fidelities. A coplanar architecture also enables us to couple multiple $\hbar$BAR devices, allowing for parallelization of MRQC operations and integration of $\hbar$BARs in general-purpose superconducting quantum circuits. These improvements would allow for further extension of operations introduced in this work, for example the implementation of quantum routing protocols\cite{xu2023systems}, thereby laying the groundwork for a hardware-efficient quantum random-access memory\cite{hann2019hardware}.

\newpage



\section*{Acknowledgements} Devices were fabricated at the FIRST cleanroom of ETH Zürich and the BRNC cleanroom of IBM Zürich. 
\paragraph*{Funding:}
YY and IK were supported by the Swiss National Science Foundation Grant no. 200021\_204073. MF and MEK were supported by the Swiss National Science Foundation Ambizione Grant no. 208886. MF and SM were supported by the Branco Weiss Fellowship – Society in Science, administered by the ETH Zurich. AO, SS, and MB were supported by the QuantERA II Program funded by the European Union's Horizon 2020
research and innovation program under grant agreement no. 101017733. RGB was supported by an ETH Grant 23-2 ETH-041. AB was supported by the Swiss National Science Foundation Grant no. CRSII--222812.
\paragraph*{Author contributions:}
YY and IK wrote experiment control sequences, performed measurements, analyzed the data, and derived theoretical models. YY, IK, MS and ME performed numerical simulations of the experiments. YY, IK, SM, SS and JE developed the code for randomized benchmarking. RGB, MEK, AO, AB and MF fabricated the device. MB fabricated the readout amplifier device. YC supervised the work. YY, IK and YC wrote the manuscript with feedback from all authors.
\paragraph*{Competing interests:} The authors declare no competing interests.
\paragraph*{Data and code availability:} All data and software are available in the manuscript or the supplementary material or are deposited at Zenodo \cite{dataset}.

\subsection*{Supplementary materials} 
Supplementary text\\
Figures S1 to S7\\
Tables S1 to S3\\
References (33-39)


\newpage


\renewcommand{\thefigure}{S\arabic{figure}}
\renewcommand{\thetable}{S\arabic{table}}
\renewcommand{\theequation}{S\arabic{equation}}
\renewcommand{\thepage}{S\arabic{page}}
\renewcommand{\thesubsection}{\Alph{subsection}}  
\setcounter{figure}{0}
\setcounter{table}{0}
\setcounter{equation}{0}
\setcounter{page}{1} 

\bibliography{MRQC}
\bibliographystyle{sciencemag}

\begin{center}
\section*{Supplementary Materials for\\ \scititle}

Yu~Yang$^{1,2,\dagger}$,
	Igor~Kladarić$^{1,2,\dagger,*}$,
	Martynas Skrabulis$^{1,2}$,
    Michael Eichenberger$^{1,2}$,
    Stefano~Marti$^{1,2}$,
    Simon Storz$^{1,2}$,
    Jonathan Esche$^{1,2}$
    Raquel García Bellés$^{1,2}$,
    Max Emanuel Kern$^{1,2}$,
    Andraz Omahen$^{1,2}$,
    Arianne Brooks$^{1,2}$,
    Marius Bild$^{1,2}$,
    Matteo Fadel$^{1,2}$,
    Yiwen~Chu$^{1,2,*}$\\ 
\small$^{1}$Department of Physics, ETH Zürich, 8093 Zürich, Switzerland \\
\small$^{2}$Quantum Center, ETH Zürich, 8093 Zürich, Switzerland \\
\small$^\ast$Corresponding author. Email: ikladaric@phys.ethz.ch; yiwen.chu@phys.ethz.ch \\
\small$^\dagger$These authors contributed equally to this work.
\end{center}

\subsubsection*{This PDF file includes:}
Supplementary text\\
Figures S1 to S7\\
Tables S1 to S3\\
References (33-39)

\newpage

\tableofcontents

\clearpage
\newpage
\subsection{System parameters}
\begin{table}[h]
\label{param_table}
\centering 

\renewcommand{\arraystretch}{0.7}
\begin{tabular}{| c | c | c |c }
\hline
\textbf{Variable} & \textbf{Parameter} & \textbf{value}\\ \hline
$g/2\pi$ & transmon-phonon coupling & $296 \text{ kHz}$ \\
$FSR/2\pi$ & phonon free spectral range & $12.6 \text{ MHz}$ \\
$\omega_r/2\pi$ & readout resonator frequency & $8.668 \text{ GHz}$ \\
\hline
\hline
$\omega_{q}/2\pi$ & transmon rest point frequency & $5.057\text{GHz}$\\
$T_{1,q}$ & transmon relaxation time & $30 \text{ $\mu$s}$ \\
$T^{E}_{2, q}$ & transmon coherence time (echo)& $23 \text{ $\mu$s}$ \\
\hline
\hline
$\omega_{p1}/2\pi$ & phonon mode 1 frequency & $5.088\text{GHz}$\\
$T_{1, p1}$ & phonon mode 1 relaxation time & $196\text{ $\mu$s}$\\
$T_{2, p1}$ & phonon mode 1 coherence time& $368\text{ $\mu$s}$ \\
\hline
$\omega_{p2}/2\pi$ & phonon mode 2 frequency & $5.076\text{GHz}$\\
$T_{1, p2}$ & phonon mode 2 relaxation time & $ 137\text{ $\mu$s}$\\
$T_{2, p2}$ & phonon mode 2 coherence time& $ 250\text{ $\mu$s}$ \\
\hline
$\omega_{p3}/2\pi$ & phonon mode 3 frequency & $5.064\text{GHz}$\\
$T_{1, p3}$ & phonon mode 3 relaxation time & $ 64\text{ $\mu$s}$\\
$T_{2, p3}$ & phonon mode 3 coherence time& $ 127\text{ $\mu$s}$ \\
\hline
\end{tabular}

\caption{\textbf{List of device properties.} The transmon frequency is controlled via the AC Stark-shift effect using a microwave drive 150 MHz detuned from the readout resonator cavity. Stronger AC Stark-shifts result in lower transmon frequencies. All transmon parameters are measured at the rest point frequency, which is also the frequency at which single-qubit gates are applied. The rest point frequency is chosen such that the transmon is off-resonant with any phonon modes and lower in frequency than all phonon modes used in the protocols. The latter choice is in order to avoid any unwanted interactions between the transmon and populated phonon modes during readout due to the dispersive effects of the readout cavity on the transmon.}
\label{simulation_params}
\end{table}

\subsection{Single-qubit gate randomized benchmarking}

For single-qubit gate randomized benchmarking, we use 24 gates from the single-qubit Clifford group\cite{barends2014superconducting}, listed in Table~\ref{gateset}. For each sequence of length $n$, the first $n-1$ gates are chosen randomly from this set, and the $n$-th gate is selected such that the transmon ideally ends in the $\ket{g}$ state. The resulting survival probability is fit to the model
\begin{equation}
    P(n) = A p^n + c,
\end{equation}
where $P(n)$ is the probability of measuring the transmon state in $\ket{g}$, $A$ and $c$ are constants determined by readout contrast, $1-p$ is the depolarization rate, and the average gate fidelity is given by $F = \tfrac{p+1}{2}$.

\begin{figure}
\centering
\includegraphics[width=16cm]{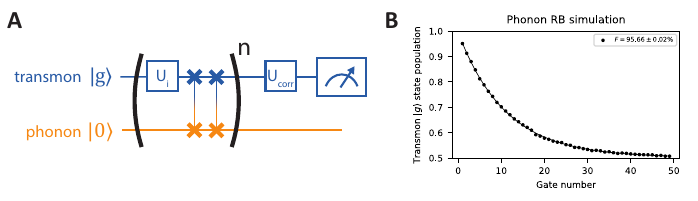}
\caption{ 
\textbf{Randomized benchmarking protocol and simulation.} 
\textbf{A}, Sequence for the single-phonon gate randomized benchmarking. $U_i$ are randomly chosen from a set of gates given in Table.~\ref{gateset}, while $U_{corr}$ is chosen such that it reverts the state of the transmon to state $\ket{g}$ in the ideal case. \textbf{B}, Simulated single-phonon gate randomized benchmarking of phonon mode 1. The obtained average Clifford gate fidelity of $F=95.66\%$ agrees well with the measured values found in the main text.
}
\label{RB_sim}
\end{figure}

Single-phonon gates additionally involve an additional pair of SWAP gates between the transmon and the phonon mode inserted after each selected transmon gate. For single-phonon randomized benchmarking, we therefore add an additional pair of SWAP gates after each transmon gate, as depicted in Fig~\ref{RB_sim}A. The phase of the single-transmon operations $U_i$ and $U_{corr}$ is calibrated so that it takes into account the phase accumulated during the SWAP operations. 
A simulation of single-phonon RB is depicted in Fig.~\ref{RB_sim}B, and its result agrees with the measured data.

\begin{table}[h!]
\centering
\caption{\textbf{Single Qubit Clifford Gates: Axis-Angle Representation.}}
\begin{tabular}{ccc}
\toprule
\textbf{Rotation Angle} & \textbf{Rotation Axis (x, y, z)} \\
\midrule
\multicolumn{3}{c}{Identity and Pauli Gates} \\
\midrule
0     & None         \\
$\pi$ & (1, 0, 0)     \\
$\pi$ & (0, 1, 0)     \\
$\pi$ & (0, 0, 1)      \\
\midrule
\multicolumn{3}{c}{Single-Axis $\pi/2$ Rotations} \\
\midrule
$\pi/2$ & $(\pm1, 0, 0)$    \\
$\pi/2$ & $(0, \pm1, 0)$    \\
$\pi/2$ & $(0, 0, \pm1)$    \\
\midrule
\multicolumn{3}{c}{$\pi$ Rotations} \\
\midrule
$\pi$ & $(\pm1, 0, 1)$    \\
$\pi$ & $(0, \pm1, 1)$    \\
$\pi$ & $(1, \pm1, 0)$    \\
\midrule
\multicolumn{3}{c}{$\pm 2\pi/3$ Rotations} \\
\midrule
$2\pi/3$  & $(\pm1, 1, 1)$   \\
$2\pi/3$  & $(\pm1, -1, 1)$  \\
$-2\pi/3$ & $(\pm1, 1, 1)$   \\
$-2\pi/3$ & $(\pm1, -1, 1)$  \\
\bottomrule
\end{tabular}
\label{gateset}
\end{table}

\subsection{Simulations}

The simulations of $C_{\phi}$ and QFT tomography protocols shown in the main text were performed using the QuTiP module for quantum information processing in Python\cite{Li2022pulselevelnoisy}. The simulations use parameters shown in Table~S1 and take into account the time-dependence of transmon drive and AC Stark-shift pulses. The transmon and phonon coherence times and their mutual coupling strengths are assumed to be independent of the transmon frequency. The transmon is modeled as a two-level system, while all phonon modes are modeled as harmonic oscillators truncated to 3 energy levels.

The state preparation and measurement protocols involve transmon pulses and SWAP operations that can either be directly simulated or implemented as ideal operations, depending on whether or not we would like to take into account SPAM errors in the simulation. The transmon state readout is assumed to be perfect and projective. The transmon state populations are obtained from diagonal components of output density matrices of the system of any simulated protocol. Experimentally, the readout of the transmon state takes $4 \, \text{$\mu s$}$, with an additional $3 \, \text{$\mu s$}$ waiting time for leakage of the readout resonator after the readout pulse. To simulate the effects of decoherence of the transmon and phonon modes during this process, in the simulations we approximate it with a $7 \, \text{$\mu s$}$ long waiting time after an ideal projective measurement.

\subsection{Tomography}

\subsubsection{Quantum state tomography}
\label{QST}
We use quantum state tomography (QST) to extract the density matrix of our system based on a set of measurements \cite{altepeter20044}.
QST relies on the fact that any n-qubit density matrix $\rho$ can be rewritten as a linear combination of tensor products of $n$ Pauli matrices $\{\mathbb{I} , X, Y, Z\}$:
\begin{equation}
\rho = \sum_{\sigma_1,\sigma_2,...,\sigma_n \in \{\mathbb{I} , X, Y, Z\}} c_{\sigma_1,\sigma_2,...,\sigma_n} \bigotimes_{i=1}^{n} \sigma_i.
\end{equation}
The coefficients $c_{\boldsymbol{\sigma}}$ can then be found by
\begin{equation}
c_{\sigma_1,\sigma_2,...,\sigma_n} = Tr\left(\left[\bigotimes_{i=1}^{n} \sigma_i\right] \rho\right),
\end{equation}
and can be obtained by measuring each qubit $i$ along the axis given by $\sigma_i$. Each Pauli operator has two eigenstates, $\ket{\psi}_{\sigma_i}$ and $\ket{\psi^{\perp}}_{\sigma_i}$, which are measured with probabilities $P(\ket{\psi}_{\sigma_i})$ and $P(\ket{\psi^{\perp}}_{\sigma_i})$. The wanted coefficient can then be calculated as
\begin{equation}
c_{\sigma_1,\sigma_2,...,\sigma_n} = \bigotimes_{i=1}^{n} \left( P(\ket{\psi}_{\sigma_i}) \pm  P(\ket{\psi^{\perp}}_{\sigma_i})\right),
\end{equation}
where the sign is determined by whether $\sigma_i = \mathbb{I}$ ($+$) or $\sigma_i \in \{X,Y,Z\}$ ($-$).

We execute quantum state tomography by applying an appropriate qubit rotation to measure the qubit along the $X, Y$, or $Z$ axis before reading out its state. 

\subsubsection{Quantum process tomography}
We characterize the fidelity of applied operations via quantum process tomography (QPT) \cite{mohseni2008quantum}. Any density matrix of an $n$-qubit system can be described in its vectorized form $\vert \rho \rangle\rangle$, containing $4^n$ elements, such that $\rho_{ij} = \vert \rho \rangle\rangle_{2^nj+i}$. Any quantum process can then be described via a superoperator $\mathcal{E}$ given as
\begin{equation}
\label{single_superop_eq}
\vert \rho \rangle\rangle_{out} = \mathcal{E} \vert \rho \rangle\rangle_{in},
\end{equation}
where $\vert \rho \rangle\rangle_{in}$ is the input and $\vert \rho \rangle\rangle_{out}$ the output vectorized density matrix.

Since $\mathcal{E}$ is a $4^n \times 4^n$ matrix, we require at least $4^n$ pairs of input and output density matrices to uniquely define the superoperator $\mathcal{E}$. We choose initial states to be tensor products of states $\{\ket{0}, \ket{1}, \ket{+}, \ket{i}\}$ for each of the $n$ qubits used, resulting in $4^n$ input combinations. We then apply the operation we wish to characterize and extract the resulting output state for each of the input states using the QST protocol described in Sec. \ref{QST}. 

The input (output) states can be grouped into a $4^n \times 4^n$ input (output) matrix $\Lambda_{in}$ ($\Lambda_{out}$) such that each of the $4^n$ columns of the matrix is given by one of the states $\vert \rho \rangle\rangle_{in}$ ($\vert \rho \rangle\rangle_{out}$). Eq.~\ref{single_superop_eq} can then be rewritten as
\begin{equation}
\label{superop_eq}
\Lambda_{out} = \mathcal{E} \Lambda_{in},
\end{equation}
allowing us to find the superoperator $\mathcal{E} = \Lambda_{out} \Lambda_{in}^{-1}$. The average gate fidelity of an applied operation $\mathcal{E}$ with a target superoperator $\mathcal{E}_{ideal}$ can then be calculated as \cite{pedersen2007fidelity}
\begin{equation}
\label{avg_gate_fid}
F_{\mathcal{E}} = \frac{2^n+Tr(\mathcal{E}\mathcal{E}_{ideal})}{2^n(2^n+1)}.
\end{equation}

Usually, our target operation is a unitary operator $U_{ideal}$, such as a $C_{\phi}$ gate. Its superoperator is then given by $\mathcal{E}_{ideal} = U_{ideal} \otimes U_{ideal}^*$. In practice, the applied operation also induces additional single-qubit rotations to each of the n qubits involved. Even if the applied operation is equivalent to the target operation up to single-qubit rotations, which can be compensated, these single-qubit rotations affect the resulting fidelity given by Eq.~\ref{avg_gate_fid}. To compensate for the single-qubit rotations, we compare the applied operation to $\tilde{U}_{ideal} = R^z_{n}(\boldsymbol{\phi}) U_{ideal}$ instead. Here 
\begin{equation}
R^z_{n}(\boldsymbol{\phi}) = \bigotimes_{i=1}^{n} R^z(\phi_i)
\end{equation}
is a tensor product of n single-qubit Z-rotations. We then optimize the set of Z-rotation phases $\boldsymbol{\phi}$ to find an ideal superoperator $\tilde{\mathcal{E}}_{ideal} = \tilde{U}_{ideal}^{} \otimes \tilde{U}_{ideal}^*$ that results in the highest fidelity given by Eq.~\ref{avg_gate_fid}. The optimal set of single-qubit rotation phases $\boldsymbol{\phi}_{optimal}$ can then be used in a subsequent round of process tomography by correcting the choice of measurement axes by the values given by $\boldsymbol{\phi}_{optimal}$.

\subsubsection{$\chi$ process matrix representation}

After obtaining the measured superoperator for quantum process we want to characterize, we visualize it using the $\chi$ process matrix representation defined by
\begin{equation}
\label{chi_formula}
\rho_{out} = \sum_{ij} \chi_{ij} E_i \rho_{in} E_j^{\dagger},
\end{equation}
where $\chi_{ij}$ are $\chi$ matrix components and $\{ E_k \}$ is a basis set of Kraus operators for which $\sum_k E_k^{\dagger} E_k = 1$. A common choice of Kraus operators is the set of tensor products of Pauli operators $\{\mathbb{I}, X, Y, Z\}$, yielding $4^n$ Kraus operators in total, making $\chi$ a $4^n \times 4^n$ matrix.

We can obtain the relationship between the $\chi$ matrix and the equivalent superoperator $\mathcal{E}$ by vectorizing Eq.~\ref{chi_formula} and using the fact that
\begin{equation}
\vert AXB \rangle\rangle = (B^T \otimes A) \vert X \rangle\rangle
\end{equation}
to find
\begin{equation}
\vert \rho \rangle\rangle_{out} = \sum_{ij} \chi_{ij} (E_j^* \otimes E_i) \vert \rho \rangle\rangle_{in}.
\end{equation}
This means that the superoperator $\mathcal{E}$ is given by
\begin{equation}
\label{superop_chi_connection}
\mathcal{E} = \sum_{ij} \chi_{ij} (E_j^* \otimes E_i).
\end{equation}
We can now vectorize $\mathcal{E} \rightarrow \vert \mathcal{E}\rangle\rangle$, $\chi \rightarrow \vert \chi\rangle\rangle$ as well as $E_i^* \otimes E_j \rightarrow \vert EE_{4^nj + i}\rangle\rangle$. If we now define a matrix $M = [\vert EE_{0}\rangle\rangle, \vert EE_{1}\rangle\rangle\, ..., \vert EE_{4^n-1}\rangle\rangle  ]$, we can rewrite Eq.~\ref{superop_chi_connection} as
\begin{equation}
\label{chi_equation}
\vert \mathcal{E}\rangle\rangle = M \vert \chi\rangle\rangle.
\end{equation}
We obtain the $\chi$ matrix by solving Eq.~\ref{chi_equation} and reconverting the vectorized result into a matrix form.

The visualizations of quantum processes in the main text are obtained by converting the measured superoperator $\mathcal{E}$ into its equivalent $\chi$ matrix representation and plotting its matrix elements. For simplicity of visualization, in Fig.~\ref{fig4} the labels of all 64 Pauli operator tensor products $\{ \mathbb{I} \otimes \mathbb{I} \otimes \mathbb{I},\,\, \mathbb{I} \otimes \mathbb{I} \otimes \mathbb{X},\,\, ...\,\, Z \otimes Z \otimes Z \}$ are indexed with values 0 to 63.

\subsubsection{Single-shot readout misassignment compensation}

Imperfect readout of the transmon results in misassignment of the transmon state and erroneous estimation of its state population statistics. If the transmon was prepared in the ground (excited) state, there is some probability $f_g$ ($f_e$) that it will be measured correctly, and a probability $1-f_g$ ($1-f_e$) that it will be misassigned to the other state. If the real populations of the transmon's ground and excited states are $p_g$ and $p_e$, respectively, then the measured populations $p_g'$ and $p_e'$ are given by
\begin{equation}
\begin{pmatrix}
f_g & 1-f_e \\
1-f_g & f_e \\
\end{pmatrix}
\begin{pmatrix}
p_g\\
p_e\\
\end{pmatrix}
=
\begin{pmatrix}
p_g'\\
p_e'\\
\end{pmatrix}.
\end{equation}
If we know the missassignment probabilities $f_g$ and $f_e$, we can reconstruct the real transmon state populations by matrix inversion
\begin{equation}
\begin{pmatrix}
p_g\\
p_e\\
\end{pmatrix}
=
\begin{pmatrix}
f_g & 1-f_e \\
1-f_g & f_e \\
\end{pmatrix}^{-1}
\begin{pmatrix}
p_g'\\
p_e'\\
\end{pmatrix},
\end{equation}
Assuming that the transmon state preparation infidelity is insignificant compared to readout infidelity, the misassignment probability $f_g$ ($f_e$) can be obtained by preparing the transmon in the ground (excited) state and measuring their populations. We then find $f_g = p_g'(\ket{0})$ and $f_e = p_e'(\ket{1})$. Our single-shot readout yields $f_g = 0.88$ and $f_e = 0.85$.

\subsection{Controlled-phase gates}

\subsubsection{Controlled-phase gate theory}
For a single phonon mode, the Hamiltonian of the system given in Eq. 1 of the main text is
\begin{equation}
H = \frac{\omega_q}{2} \sigma_z + \omega_p a^{\dagger} a + g \left( \sigma_+ a + \sigma_- a^{\dagger}\right).
\end{equation}
In the rotating frame of the phonon mode, this becomes
\begin{equation}
H = \frac{\Delta}{2} \sigma_z + g \left( \sigma_+ a + \sigma_- a^{\dagger}\right),
\end{equation}
or in matrix form
\begin{equation}
\begin{pmatrix}
-\frac{\Delta}{2} & 0&0 &0 & 0& \\
0& \frac{\Delta}{2} & g & 0&0 & \hdots\\
0& g & -\frac{\Delta}{2} &0 &0 & \\
0& 0& 0& \frac{\Delta}{2} & g \sqrt{2} & \\
0&0 & 0& g \sqrt{2} & -\frac{\Delta}{2} & \\
& \vdots & & & & \ddots
\end{pmatrix}
\begin{matrix}
     \ket{g0}\\
     \ket{e0}\\
     \ket{g1}\\
     \ket{e1}\\
     \ket{g2}\\
     \vdots
\end{matrix} \;,
\end{equation}
We can focus on individual blocks of the Hamiltonian
\begin{equation}
H_n/\hbar = \begin{pmatrix}
\frac{\Delta}{2} & g \sqrt{n} \\
g \sqrt{n} & -\frac{\Delta}{2} \\
\end{pmatrix}
\begin{matrix}
    \ket{e, n-1}\\
    \ket{g, n}
\end{matrix}\,.
\end{equation}
The eigenvalues of this Hamiltonian are given as 
\begin{equation}
\lambda_{n,\pm} = \pm \frac{\omega_n}{2},
\end{equation}
where  we introduced $\omega_n = \sqrt{\Delta^2+4g^2n}$. \newline
The eigenvectors of this Hamiltonian can be expressed as
\begin{equation}
\label{new_eigenbasis}
\ket{\lambda_{n,\pm}} = a_{n,\pm} \ket{e, n-1} + b_{n,\pm} \ket{g, n},
\end{equation}
with 

\begin{equation}
\label{ab_expressions}
\begin{split}
a_{n,\pm} = \frac{\lambda_{n,\pm} + \frac{\Delta}{2}}{\sqrt{g^2n+\lambda^2_{n,\pm}}}, \\
b_{n,\pm} = \frac{g\sqrt{n}}{\sqrt{g^2n+\lambda^2_{n,\pm}}}.
\end{split}
\end{equation}
We can then rewrite the original eigenbasis as

\begin{equation}
\begin{split}
\ket{e, n-1} = \frac{b_-}{a_+b_- - a_-b_+}\ket{\lambda_+} + \frac{b_+}{a_+b_- - a_-b_+}\ket{\lambda_-}, \\
\ket{g, n} = -\frac{a_-}{a_+b_- - a_-b_+}\ket{\lambda_+} + \frac{a_+}{a_+b_- - a_-b_+}\ket{\lambda_-}.
\end{split}
\end{equation}
The unitary evolution $U_n(t) = e^{-iH_n t/ \hbar}$ of states $\ket{\lambda_+}$ and $\ket{\lambda_-}$ under the Hamiltonian $H_n$ results in accumulated phases $-i\lambda_+ t$ and $-i\lambda_- t$, respectively. Rewriting the evolved states in terms of the original eigenbasis using \ref{new_eigenbasis} allows us to obtain the expression

\begin{equation}
U_n(t) = \frac{1}{a_+b_- - a_-b_+}\begin{pmatrix}
a_+b_-e^{-i\lambda_+t} - a_-b_+e^{-i\lambda_-t} & -a_+a_-\left(e^{-i\lambda_+t}-e^{-i\lambda_-t}\right) \\
b_+b_-\left(e^{-i\lambda_+t}-e^{-i\lambda_-t}\right) & a_+b_-e^{-i\lambda_-t} - a_-b_+e^{-i\lambda_+t} \\
\end{pmatrix}
\begin{matrix}
    \ket{e, n-1}\\
    \ket{g, n}
\end{matrix}\,.
\end{equation}
Inserting expressions given by \ref{ab_expressions} allows us to finally obtain the expression for the unitary describing the off-resonant transmon-phonon interaction as

\begin{equation}
\label{unitary_general}
U_n(t) = e^{-i\frac{\Delta}{2}t}\begin{pmatrix}
\cos\left(\frac{\omega_n}{2}t\right) - i\frac{\Delta}{\omega_n} \sin\left(\frac{\omega_n}{2}t\right) & -2i\frac{g}{\omega_n} \sin\left(\frac{\omega_n}{2}t\right) \\
-2i\frac{g}{\omega_n} \sin\left(\frac{\omega_n}{2}t\right) & \cos\left(\frac{\omega_n}{2}t\right) + i\frac{\Delta}{\omega_n} \sin\left(\frac{\omega_n}{2}t\right) \\
\end{pmatrix}.
\end{equation}
For simplicity, we can rewrite the unitary as

\begin{equation}
U_n(t) = \begin{pmatrix}
x_n & y_n \\
y_n & x_n^* \\
\end{pmatrix}.
\end{equation}
The full unitary operation is then given as
\begin{equation}
\label{full_unitary}
U(t) =  e^{-i\frac{\Delta}{2}t}
\begin{pmatrix}
1 & 0&0 &0 & 0& \\
0& x1 & y1 & 0&0 & \hdots\\
0& y1 & x1^* &0 &0 & \\
0& 0& 0& x2 & y2 & \\
0&0 & 0& y2 & x2^* & \\
& \vdots & & & & \ddots
\end{pmatrix}
\begin{matrix}
     \ket{g0}\\
     \ket{e0}\\
     \ket{g1}\\
     \ket{e1}\\
     \ket{g2}\\
     \vdots
\end{matrix} \;,
\end{equation}

Our implementation of the $C_{\phi}$ gate involves two such off-resonant interactions separated by a Z-rotation $R(\theta)$ of the transmon producing $U(t)R(\theta)U(t)$. $R(\theta)$ is composed of blocks

\begin{equation}
R_n(\theta) = \begin{pmatrix}
e^{-i\theta} & 0 \\
0 & 1 \\
\end{pmatrix}
\begin{matrix}
    \ket{e, n-1}\\
    \ket{g, n}
\end{matrix}\,.
\end{equation}
A single block of the total unitary operation is then found to be
\begin{equation}
U_n(t)R_n(\theta)U_n(t) = e^{-i\Delta t}\begin{pmatrix}
x_n^2 e^{-i\theta} + y_n^2 & y_n\left(x_n e^{-i\theta}+x_n^*\right) \\
y_n\left(x_n e^{-i\theta}+x_n^*\right) & y_n^2 e^{-i\theta} + x_n^2 \\
\end{pmatrix}.
\end{equation}

The $C_{\phi}$ gate is diagonal, so we require $y_n\left(x_n e^{-i\theta}+x_n^*\right) = 0$ for n=1,2. Since $y_n = -2i\frac{g}{\omega_n} \sin\left(\frac{\omega_n}{2}t\right)$, it is not possible to simultaneously achieve $y_1 = y_2 = 0$ due to $\omega_1 \neq \omega_2$. Similarly $x_n e^{-i\theta}+x_n^* = 0$ for $\theta = \pm \pi - 2\arctan\left( \frac{\Delta}{\omega_n} \tan \left( \frac{\omega_n}{2}t \right)\right)$ and therefore also cannot be $0$ for both $n=1$ and $n=2$ simultaneously. 

To achieve a diagonal unitary operation, we then require $y_2 = 0$ and $x_1 e^{-i\theta}+x_1^* = 0$, or vice-versa. Because we want the operation to take as little time as possible to reduce decay and decoherence of the system during operation, we choose $y_2=0$, which is achieved for
\begin{equation}
\label{t_int_expression}
t_{int} = \frac{2\pi}{\omega_2} = \frac{2\pi}{\sqrt{\Delta^2+8g^2}}
\end{equation}
This results in the required $\theta$ to be
\begin{equation}
\label{theta_expresssiom}
\theta = \pm \pi - 2\arctan\left( \frac{\Delta}{\omega_1} \tan \left( \frac{\omega_1}{2}t_{int} \right)\right)
\end{equation}
Plugging these results into Eq.~\ref{unitary_general} and looking only at the lowest four energy levels, we obtain
\begin{equation}
U(t)R(\theta)U(t) = 
\begin{pmatrix}
1 & 0&0 &0 \\
0& -e^{-\Delta t_{int}} & 0 & 0\\
0& 0 & -e^{-\Delta t_{int}-\theta} &0\\
0& 0& 0& e^{-\Delta t_{int} - \theta}
\end{pmatrix}
\begin{matrix}
     \ket{g0}\\
     \ket{e0}\\
     \ket{g1}\\
     \ket{e1}\\
\end{matrix} \;,
\end{equation}
or more precisely, the accumulated phases for each initial state are $\phi_{g0}=0$, $\phi_{e0}=-\Delta t_{int} \pm \pi$, $\phi_{g1}=-\Delta t_{int} -\theta \pm \pi$, $\phi_{e1}=-\Delta t_{int} -\theta$.
The $C_{\phi}$ controlled phase $\phi$ is then given as 
\begin{equation}
\phi = \phi_{e1} - \phi_{e0} - \phi_{g1} = \Delta t_{int} \pm 2\pi
\end{equation}
To obtain the wanted detuning $\Delta$ given a target controlled phase $\phi$, we insert the expression for $t_{int}$ found in \ref{t_int_expression}. This leads to
\begin{equation}
\frac{\Delta}{\sqrt{\Delta^2+8g^2}} = \frac{\phi}{2\pi} \pm 1.
\end{equation}
Considering that $\frac{\Delta}{\sqrt{\Delta^2+8g^2}} \in [-1,1]$, it is necessary that
\begin{equation}
\label{delta_eq}
\frac{\Delta}{\sqrt{\Delta^2+8g^2}} = \frac{\phi}{2\pi} - sgn(\phi),
\end{equation}
which also implies that $sgn(\Delta) = -sgn(\phi)$.

Solving \ref{delta_eq} for $\Delta$, we find

\begin{equation}
\label{Delta_expression}
\Delta = -sgn(\phi) \frac{2g\sqrt{2}}{\sqrt{(\frac{2\pi}{\phi- sgn(\phi)\cdot2\pi})^2-1}},
\end{equation}

Together, equations \ref{t_int_expression}, \ref{theta_expresssiom} and \ref{Delta_expression} give all the necessary parameters to implement the operation $C_{\phi}$ for an arbitrary controlled phase $\phi$. The expressions are plotted in Fig.~\ref{fig2} of the main text.

\subsubsection{Controlled-phase gate calibration}
The implemented controlled-phase gate depends on three variables -- the interaction time $t_{int}$, the detuning $\Delta$, and the Z-rotation phase $\theta$. While the interaction time $t_{int}$ can be directly set, the other two parameters need to be calibrated through measurements.

The transmon-phonon detuning $\Delta$ is calibrated by measuring the frequency difference between the transmon and the phonon mode. The phonon mode frequency is measured using the Ramsey sequence protocol in the following way \cite{chu2017quantum}: a $\pi/2$-pulse is applied to the transmon at the rest point frequency and the state is then swapped into the phonon mode. After a waiting time $t$, the state is swapped back into the transmon and another $\pi/2$-pulse is applied to the transmon. Measuring the transmon population for different waiting times $t$ results in oscillations in the transmon population with frequency equal to the difference between the frequency of the phonon mode and the rest point frequency of the qubit, which can then be extracted by fitting. We then measure the transmon frequency using spectroscopy. Due to the transmon-phonon dispersive shift, the measured transmon-phonon detuning $\Delta'$ is given by $\Delta' = \sqrt{\Delta^2+4g^2}$, where $g$ is the coupling strength between the two. This allows us to find the intrinsic transmon-phonon detuning $\Delta$. We then tune the transmon frequency using the AC Stark-shift such that $\Delta$ is set to the target value given by Eq.~\ref{Delta_expression}.

\begin{figure}
\centering
\includegraphics[width=16cm]{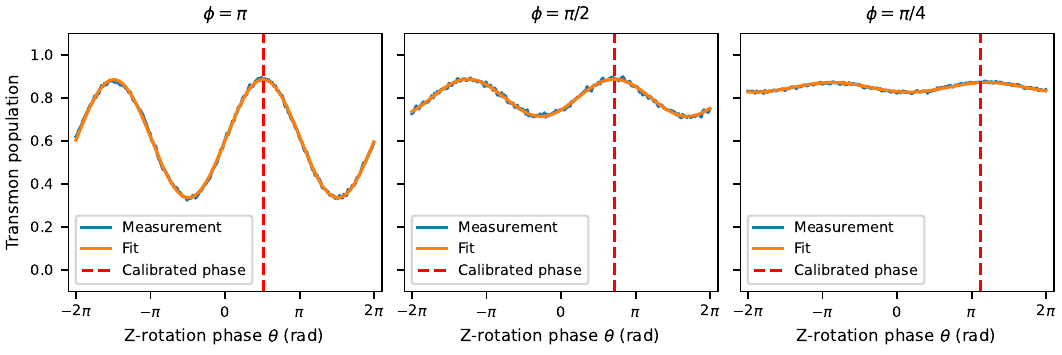}
\caption{ 
\textbf{$C_{\phi}$ Z-rotation phase calibration.} 
Calibration of the $C_{\phi}$ Z-rotation phases $\theta$ for $\phi = \pi$ (left), $\phi = \pi/2$ (middle) and $\phi = \pi/4$ (right).
}
\label{Cphase_calib_fig}
\end{figure}

Even though a theoretical expression for the necessary Z-rotation phase $\theta$ is given by \ref{theta_expresssiom}, the tuning of the transmon frequency during the off-resonant transmon-phonon interactions induces additional phase accumulation in the transmon state that needs to be taken into account during the Z-rotation operation. Therefore, the phase $\theta$ is calibrated experimentally. This is done by initializing the system in the $\ket{e0}$ state and applying the controlled-phase gate protocol with a variable Z-rotation phase $\theta$. Considering that the $C_{\phi}$ gate should result only in phase accumulation and no population exchange between the transmon and the phonon mode, the final state should again be measured in $\ket{e0}$. By sweeping the Z-rotation phase $\theta$, we observe oscillations in the final transmon $\ket{e}$ state population. We then choose the optimal phase $\theta$ to be the one for which this population reaches the maximum value. Examples of $C_{\pi}, C_{\pi/2}$ and $C_{\pi/4}$ calibration curves can be found in Fig.~\ref{Cphase_calib_fig}

\subsection{Quantum Fourier transform}

\subsubsection{Quantum Fourier transform protocol theory}

This section briefly summarizes the quantum Fourier transform protocol, closely following a more detailed introduction that can be found in \cite{nielsen2010quantum}. It also describes the adaptation of the the protocol to the MRQC platform. 

QFT is defined to act on basis states $\ket{x}$ in the following way:
\begin{equation}
\label{QFT_def}
U_{QFT} \ket{x} = \frac{1}{\sqrt{N}}\sum_{y=0}^{N-1}e^{\frac{2 \pi i x y}{N}} \ket{y}.
\end{equation}
Consequently, the operation acts on an arbitrary state as
\begin{equation}
U_{QFT} \sum_{x=0}^{N-1}\psi_x \ket{x} = \sum_{y=0}^{N-1}\psi_y \ket{y},
\end{equation}
with $\psi_y$ given by the discrete Fourier transform of $\psi_x$:
\begin{equation}
\psi_y = \frac{1}{\sqrt{N}}\sum_{x=0}^{N-1} \psi_x e^{\frac{2 \pi i x y}{N}}.
\end{equation}
If we represent the state $\ket{x}$ in binary form $\ket{x} = \ket{x_nx_{n-1}...x_1}$, with $x_n$ being the most significant bit, Eq.~\ref{QFT_def} can be rewritten as
\begin{equation}
\label{QFT_def2}
U_{QFT} \ket{x_nx_{n-1}...x_1} = \frac{\left( \ket{0} + e^{2\pi i 0.x_1}\ket{1}\right) \left( \ket{0} + e^{2\pi i 0.x_2x_1}\ket{1} \right) ... \left( \ket{0} + e^{2\pi i 0.x_nx_{n-1}...x_1}\ket{1}\right)}{\sqrt{2^n}},
\end{equation}
where $0.x_lx_{l+1}...x_{l+m}$ represents a binary fraction $x_l/2 + x_{l+1}/4 + ... + x_m/2^{m+1}$.

This allows us to formulate an algorithm for the implementation of a QFT protocol using Hadamard gates and controlled-rotations 
\begin{equation}
    R_k =\begin{pmatrix}
        1&0\\
        0&e^{\frac{2\pi i}{2^k}}
    \end{pmatrix},
\end{equation}
which are equivalent to $C_{\phi}$ gates.

\begin{figure}
\centering
\includegraphics[width=16cm]{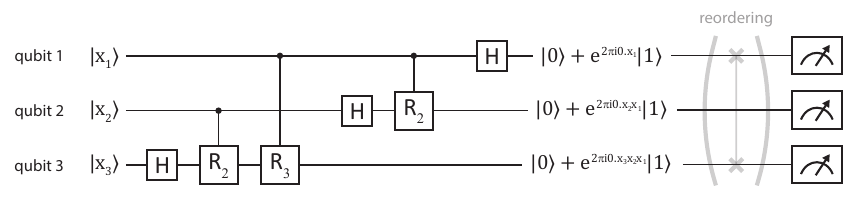}
\caption{ 
\textbf{Gate decomposition of the QFT protocol.}
A $QFT_3$ sequence on a 3-qubit system. In MRQC, each qubit is implemented as a single phonon mode, while Hadamard gates and controlled-phase operations are applied via a transmon qubit.
}
\label{QFT_general_fig}
\end{figure}

The algorithm is depicted in Fig.~\ref{QFT_general_fig}. Note that the output state of the algorithm has a reversed order of bits compared to the target state of Eq.~\ref{QFT_def2}, which either needs to be compensated by additional SWAP operations, or needs to be taken into consideration for subsequent operations and measurements, as we have done in our work. The complexity of the algorithm is therefore $\Theta(n^2)$.

In our system, single- and two-qubit gates can only be applied on the transmon, or between the transmon and a phonon mode, respectively. The algorithm then needs to be adjusted such that swap operations between a phonon mode and the transmon are added before and after applying a block of operations that involve that mode. Since the number of additional necessary SWAP operations necessary for this adjustment is $O(n)$, this does not affect the complexity of the algorithm.

\subsubsection{Quantum Fourier transform calibration}
The QFT sequence consists of $C_{\phi}$ and Hadamard gates, all of which need to be calibrated for accurate QFT protocol implementation. While the $C_{\phi}$ gates can be independently calibrated, the Hadamard gates are defined relative to a particular phase reference. Specifically, he Hadamard gate is decomposed into
\begin{equation}
H = R_y(-\frac{\pi}{2}) \cdot R_x(\pi),
\end{equation}
where x- and y- axes are in the rotating frame of the mode they are applied to, which, in our case, is the rest-point frequency of the transmon. That is, for example, if the transmon is in $\ket{+}$, we expect a Hadamard gate to map it to the state $\ket{0}$ and vice-versa, regardless of where in the sequence the gate is applied. However, since the phases of states stored in the phonon modes evolve at different rates due to differing frequencies, the phases of the Hadamard pulses need to be adjusted within each sequence to take this effect into account. In other words, we need to apply operations $R_{\phi}(\pi)$ and $R_{\phi+\pi/2}(\frac{\pi}{2})$, with the freedom to choose the orientation of the operations given by angle $\phi$. The necessary phase adjustments can in principle be calculated by knowing the frequency of the phonon modes, the duration of storage of states in each of the phonon modes, as well as additional phase accumulation during the application of $C_{\phi}$ gates, all of which is accessible through calibrations described in earlier sections. 

\begin{figure}
\centering
\includegraphics[width=16cm]{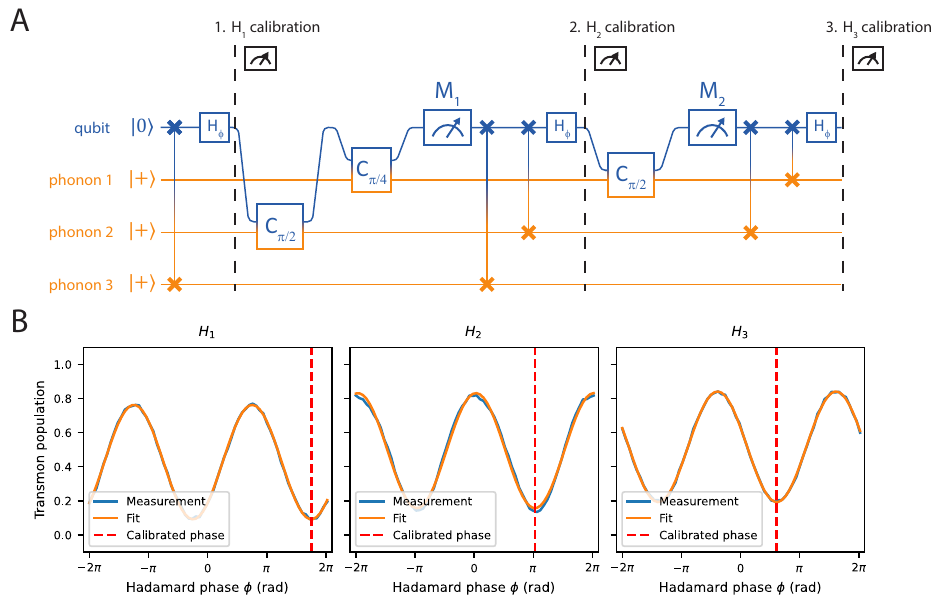}
\caption{
\textbf{Hadamard gate calibration for the QFT protocol.}
\textbf{A}, Sequence for calibrating the Hadamard gate phases of the $QFT_3$ sequence. \textbf{B}, Measured Hadamard gate calibration plots. We chose phases that result in the least transmon population.
}
\label{QFT_calibration_fig}
\end{figure}

However, for the purpose of optimizing the fidelity of the applied QFT algorithm, we instead calibrate the individual Hadamard gates experimentally in the context of the whole sequence. To do so, we apply a similar sequence to the QFT sequence shown in Fig.~\ref{QFT_calibration_fig} after having calibrated all required $C_{\phi}$ gates. Firstly, we calibrate the first Hadamard gate by initializing the phonon modes in the $\ket{+}\ket{+}\ket{+}$ state, then applying the sequence up to point 1, and sweeping the phase $\phi_1$. We choose the phase $\phi_1$ at which the measured $\ket{e}$ state population reaches the minimum, since we expect the Hadamard gate to map state $\ket{+}$ to $\ket{0}$. We then apply the protocol up to point 2, and finally to point 3 to calibrate the remaining two Hadamard gates in the same manner. The calibrated phases $\phi_1, \phi_2$ and $\phi_3$ are calibrated only once and are then used in the full QFT sequence characterization.

\subsubsection{Quantum Fourier transform error budget}

To estimate the sources of errors of the $QFT_3$ operation, we simulate the $QFT_3$ tomography protocol with different system parameters and choices of real or ideal state preparation and measurement protocols, and compare them to the measured fidelity. We consider 5 cases:

\begin{enumerate}
  \item \textbf{No SPAM error.}  We omit the errors caused by SPAM operations.  A measurement is treated as an instantaneous and ideal projective readout of the transmon state.  The process fidelity is
  \( F = 80.2\%.\)

  \item \textbf{State-preparation error only.}  We include the state-preparation errors but retain ideal measurements.  This yields
  \(F = 74.2\%.\)

  \item \textbf{Measurement error only.}  We simulate ideal state preparation, but insert an idle delay after each projective measurement to simulate system decay during readout and model the impact of the measurement. This yields
  \(F = 73.6\%.\)
  
  \item \textbf{Full SPAM error.}  We include both state preparation and measurement errors. The simulated fidelity drops to
  \(F = 68.9\%.\)
    
  \item \textbf{Measured error} We implement the protocol experimentally and measure the fidelity to obtain \(F = 54.8\%.\)
\end{enumerate}
These results allow us to estimate the following error budget for the $QFT_3$ operation:
\begin{itemize}
  \item QFT circuit operations (gates and SWAPs) contribute approximately 20\% infidelity.
  \item State-preparation errors account for about 6\% infidelity.
  \item Idle-mode decay during readout contribute the remaining 6\% infidelity.
  \item additional sources of error unaccounted by the simulations contribute approximately 14\% infidelity.
\end{itemize}
We additionally estimate the errors caused by phonon and transmon decay and decoherence:
\begin{enumerate}

 \item \textbf{Full SPAM error, but infinite phonon coherence.} We include both SPAM errors, but assume an infinite phonon coherence time to exclude the effects of phonon decay and decoherence. The simulated fidelity increases to \( F = 77.3\%  \) compared to case 4 above.
  
  \item \textbf{Full SPAM error, but infinite qubit coherence.} We include both SPAM errors, but assume an infinite qubit coherence time to exclude the effects of qubit decay and decoherence. The simulated fidelity increases to \( F = 83.4\%  \) compared to case 4 above.
\end{enumerate}
These results show that phonon decay and decoherence contributes approximately 9\%, while transmon decay and decoherence approximately 15\% infidelity, showing that our system is currently predominantly limited by transmon coherence properties.

\subsection{Quantum period finding}

\subsubsection{Quantum period finding theory}
The quantum period finding algorithm allows us to find the period $r$ of a function $f(x)$ implemented via an oracle $O_f$. The algorithm involves n data qubits and one output qubit, and can be decomposed into five parts - state preparation, oracle application, QFT application, measurement and classical post-processing. As an example, we will be using n=3 data qubits.

During the state preparation phase, we prepare each of the data qubits in the $\ket{+}$ state, with the output qubit remaining in the ground state.
\begin{equation}
\ket{\psi} = \ket{+}_3 \ket{+}_2 \ket{+}_1 \ket{0}_o = \frac{1}{\sqrt{N}}\sum_{x=0}^{N-1}\ket{x}_{d}\ket{0}_o,
\end{equation}
where x is an integer representation of the binary number stored in the data qubits, e.g. $\ket{4}_d = \ket{1}_3\ket{0}_2\ket{0}_1$.

In the second phase we apply the oracle $O_f$ that writes $f(x)$ into the output qubit
\begin{equation}
\label{before_QFT}
\ket{\psi} = \frac{1}{\sqrt{N}}\sum_{x=0}^{N-1}\ket{x}_{d}\ket{f(x)}_o.
\end{equation}
Next, we apply the QFT algorithm to the obtained state stored in the data qubits, such that we obtain
\begin{align}
\label{After_QFT}
\ket{\psi} &= \frac{1}{N}\sum_{x=0}^{N-1}\sum_{y=0}^{N-1}e^{\frac{2\pi i x y}{N}}\ket{y}_{d}\ket{f(x)}_o = \\
&= \frac{1}{N}\sum_{v=0,1}\sum_{\substack{0 \le x \le N-1 \\ f(x) = v}}e^{\frac{2\pi i x y}{N}}\ket{y}_{d}\ket{v}_o \\
&= \sum_{y=0}^{N-1}\ket{y}_d \left( \frac{1}{N} \sum_{\substack{0 \le x \le N-1 \\ f(x) = 0}}e^{\frac{2 \pi i x y}{N}} \ket{0}_o +  \frac{1}{N} \sum_{\substack{0 \le x \le N-1 \\ f(x) = 1}}e^{\frac{2 \pi i x y}{N}} \ket{1}_o\right).
\end{align}
Measuring the state of the data qubits results in the probability of measuring $\ket{y}_d$
\begin{equation}
\label{QPF_theory_prediction}
P(y) = \frac{1}{N^2} \left( \left|\sum_{\substack{0 \le x \le N-1 \\ f(x) = 0}}e^{\frac{2\pi i x y}{N}}\right|^2 + \left|\sum_{\substack{0 \le x \le N-1 \\ f(x) = 1}}e^{\frac{2\pi i x y}{N}}\right|^2 \right).
\end{equation}
If we assume that $f(x)$ is $r$-periodic, such that $f(x) = f(x+kr)$, we can rewrite
\begin{align}
\sum_{\substack{0 \le x \le N-1 \\ f(x) = v}}e^{\frac{2\pi i x y}{N}} &= \sum_{\substack{0 \le x \le r-1 \\ f(x) = v}} \sum_{k=0}^{N/r-1}e^{\frac{2\pi i (x+kr) y}{N}} = \\
&= \sum_{\substack{0 \le x \le r-1 \\ f(x) = v}} e^{\frac{2\pi i x y}{N}} \left(\sum_{k=0}^{N/r-1}e^{\frac{2\pi i k y}{N/r}}\right).
\end{align}


The expression in the parentheses is equal to $\frac{N}{r}$ when $y = l\frac{N}{r}$, with $l \in [0,r-1]$, and 0 otherwise. Equation \ref{QPF_theory_prediction} therefore simplifies to


\begin{equation}
\label{QPF_theory_prediction_r}
P(y) =
\begin{cases}
0, & y \neq l\frac{N}{r}, \, l \in [0,r-1] \\
\frac{1}{r^2} \left( \left|\sum_{\substack{0 \le x \le r-1 \\ f(x) = 0}}e^{\frac{2\pi i xy}{N}}\right|^2 + \left|\sum_{\substack{0 \le x \le r-1 \\ f(x) = 1}}e^{\frac{2\pi i xy}{N}}\right|^2 \right), & y = l\frac{N}{r}, \, l \in [0,r-1],
\end{cases}
\end{equation}
This means that we will observe peaks in the population of states that are integer multiples of $\frac{N}{r}$, while the others are ideally 0. To extract the period $r$, we can determine the peak locations and calculate their greatest common divisor $s$. The period of the applied function $f(x)$ is then given as $r = \frac{N}{s}$.

\subsubsection{Quantum period finding calibration}

\begin{figure}
\centering
\includegraphics[width=16cm]{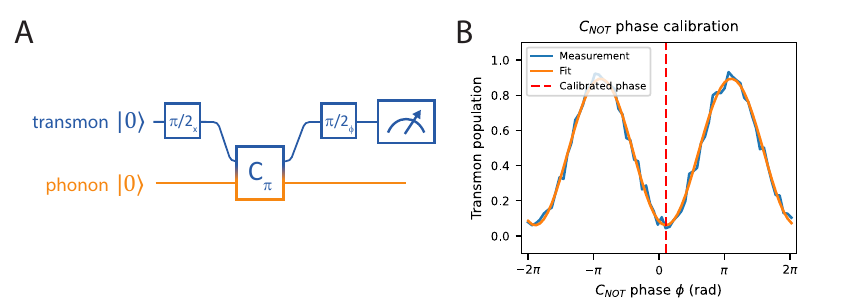}
\caption{ 
\textbf{CNOT gate sequence and calibration.}
\textbf{A}, Decomposition of the CNOT gate into $\pi/2$-pulses and a $C_{\pi}$ gate. The final transmon measurement is used for phase calibration of the second pulse. \textbf{B}, Measured calibration data for the second $\pi/2$-pulse. We choose the phase for which the transmon is measured back in the ground state.
}
\label{QPF_calibration_fig}
\end{figure}

Calibration of the QPF protocol requires calibration of controlled-phase and Hadamard gates, similarly to the QFT protocol, as well as calibration of the operation that implements the oracle $O(f(x))$. In the case of the 2-periodic function from the main text, the oracle can be implemented via a CNOT gate between the least significant phonon mode and the transmon. Considering that, after state preparation, the transmon is initialized in the ground state, we can implement the CNOT gate via a $C_{\pi}$ gate and $\pi/2$-rotations:
\begin{equation}
C_{NOT} = R_y(-\frac{\pi}{2}) \cdot C_{\pi} \cdot R_y(\frac{\pi}{2}),
\end{equation}
as depicted in Fig.~\ref{QPF_calibration_fig}A. However, since the $C_{\pi}$ gate involves shifting the transmon frequency, the phase of the second $\pi/2$-pulse needs to be calibrated. This calibration is done by preparing both the phonon mode and the transmon in the ground state and measuring the final transmon state for different phases of the second pulse. The phase for which the transmon returns to the ground state is chosen. An example of such a calibration is shown in Fig.~\ref{QPF_calibration_fig}B.

The remainder of the protocol is calibrated sequentially in a manner similar to the QFT protocol calibration described above.

\subsubsection{Quantum period finding -- other oracles}

\begin{figure}
\centering
\includegraphics[width=16cm]{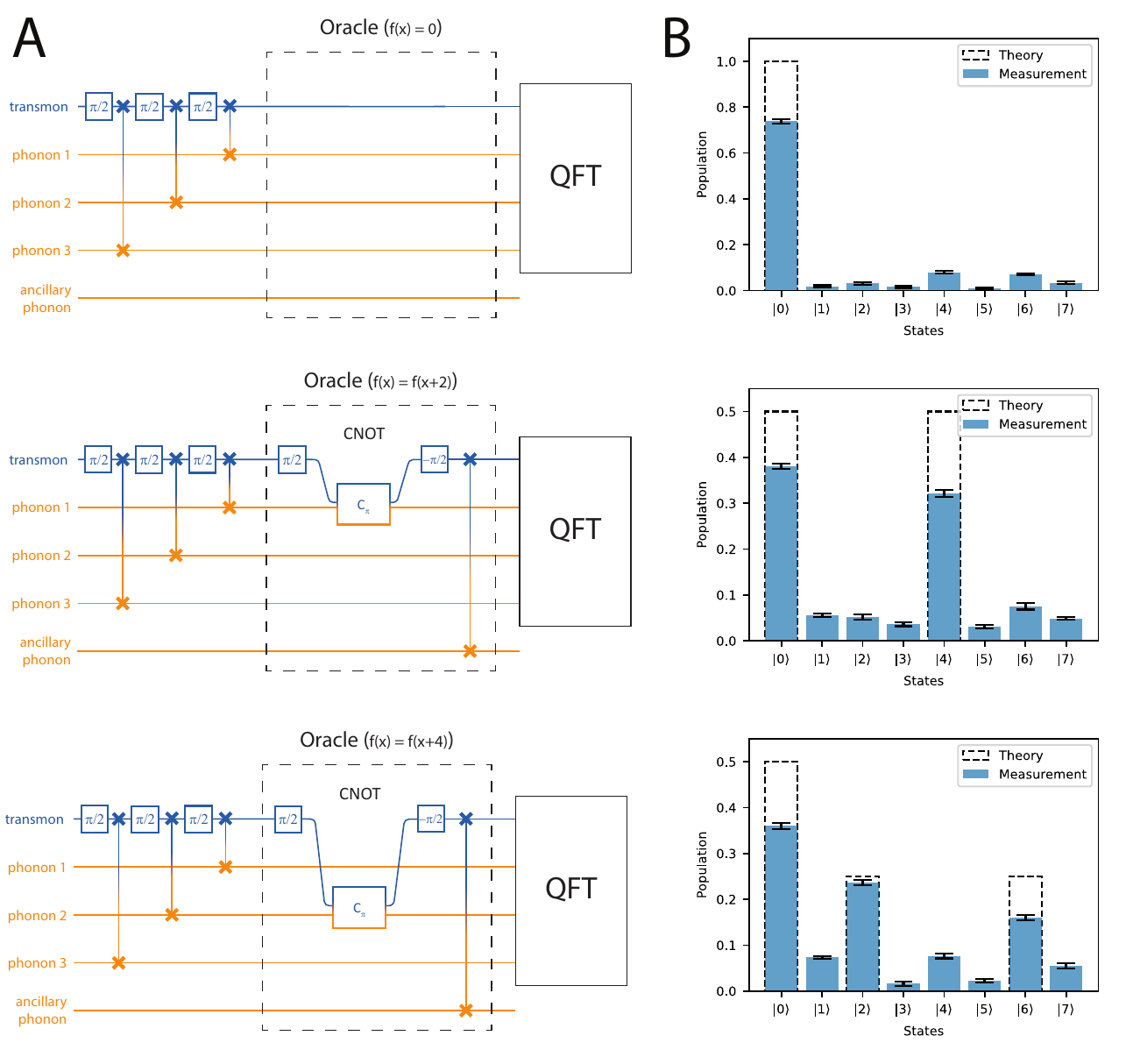}
\caption{ 
\textbf{QPF sequences and measurement results.}
\textbf{A}, Quantum period finding sequences implemented for r-periodic functions with $r=1$ (top), $r=2$ (middle) and $r=4$ (bottom) and
\textbf{B}, their corresponding measured phonon state populations.
}
\label{QPF_all_results_fig}
\end{figure}

Implementing oracles of functions with varying periods $r$ results in differing output populations. For convenience of the experiment, we assume that the period of a function $r$ needs to be a divisor of the dimension of the system $N = 2^n$. In our case, $N=8$, meaning that the period can attain values $r=1,2,4,8$. We implement oracles of functions with $r=1,2,4$ as depicted in Fig.~\ref{QPF_all_results_fig}A, which result in functions defined in Table~\ref{Function_table}.

Using Eq.~\ref{QPF_theory_prediction} or by simulating an ideal QPF protocol, we obtain populations found in Fig.~\ref{QPF_all_results_fig}B, to which we can compare experimentally measured populations shown in the same figure.

\begin{table}[h]
\centering 

\renewcommand{\arraystretch}{0.7}

\begin{tabular}{| c | c | c | c | c | c | c | c | c |}
\hline
\textbf{Period} & f(0) & f(1) & f(2) & f(3) & f(4) & f(5) & f(6) & f(7) \\ \hline \hline

\textbf{r=1} & 0 & 0 & 0 & 0 & 0 & 0 & 0 & 0 \\
\hline

\textbf{r=2} & 0 & 1 & 0 & 1 & 0 & 1 & 0 & 1 \\
\hline

\textbf{r=4} & 0 & 0 & 1 & 1 & 0 & 0 & 1 & 1 \\
\hline

\end{tabular}
\caption{\textbf{Implemented r-periodic functions f(x).}}
\label{Function_table}
\end{table}

\newpage

\subsubsection{Quantum period finding analysis}

Determining the period $r$ of a function $f(x)$ using the QPF algorithm relies on the ability to determine peaks in the state populations of the data qubits after the application of the algorithm. In theory, peaks correspond to all non-zero population values. However, noise, population leakage, gate parameter calibration imperfection, and readout infidelity result in non-zero values in all possible output states. Therefore, the most probable peaks need to be detected that can then be used for subsequent post-processing and determination of the period $r$. To classify which states most likely correspond to zero or non-zero populations, we assume a two-component mixture model of probability distributions corresponding to the two groups, and find the most likely distributions using the expectation-maximization algorithm (EM)\cite{dempster1977maximum}.

\begin{figure}
\centering
\includegraphics[width=16cm]{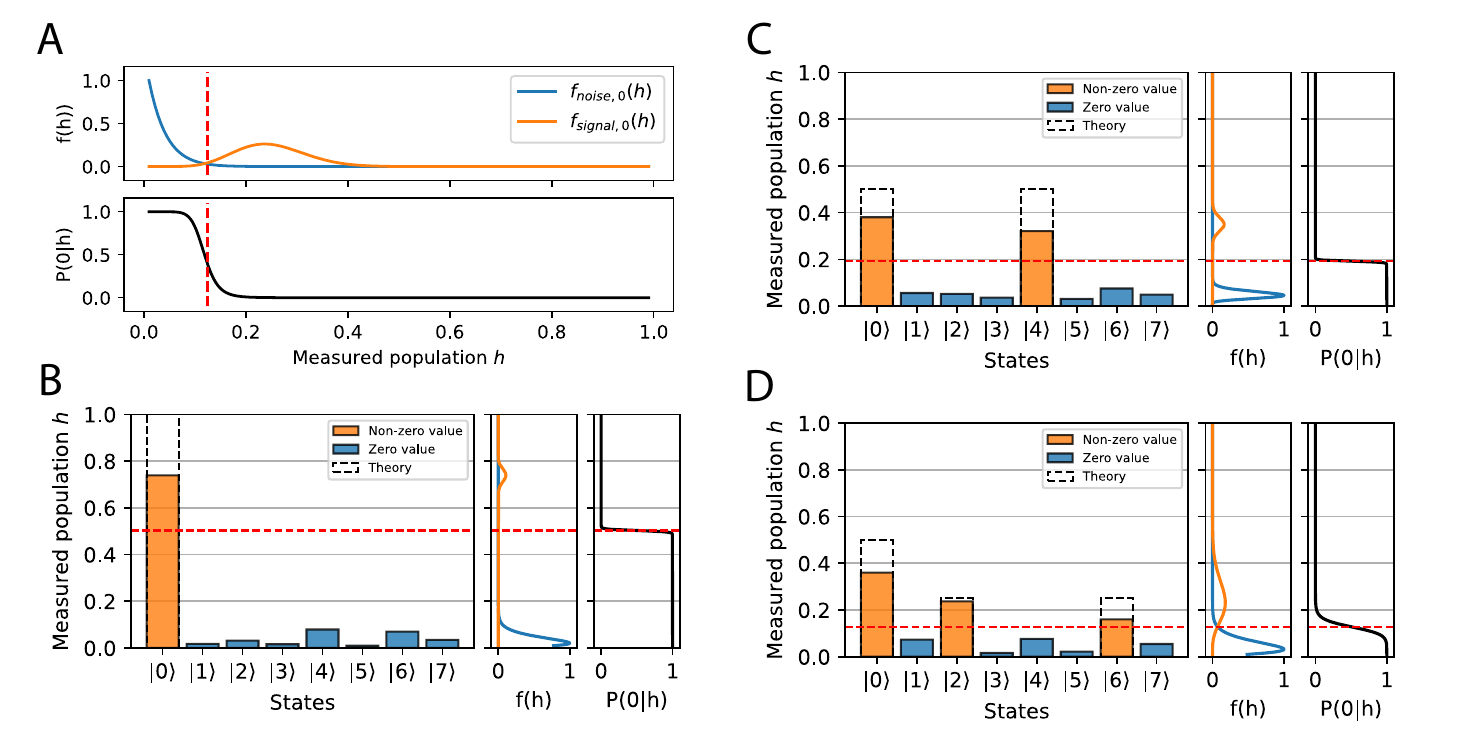}
\caption{ 
\textbf{QPF peak classification and period determination.}
\textbf{A}, Initial guesses for $f_{noise}(h)$ and $f_{signal}(h)$ (top), and the resulting zero-value probability $P(0|h)$ given a measured population h. 
\textbf{B, C, D}, classification of zero and non-zero values for $r=1,2,4$, respectively. (left) Measured populations $h$ (colored) and their comparison to the theoretical prediction (dashed); (middle) final $f(h)$ probability distributions, and (right) the resulting $P(0|h)$ on which the classification is based on. The red dashed line represents the measured population $h$ for which $P(0|h)=0.5$.
}
\label{QPF_MM}
\end{figure}

Measuring the population of states $\ket{i}$ will result with values $h_i$. States with the highest likelihood to have zero population will then attain values determined only by the noise probability distribution $f_{noise}(h)$, while those with non-zero population will follow a different probability distribution $f_{signal}(h)$. Since $h \in [0,1]$, we choose to model our probability distributions as $\beta$-functions\cite{berger2013statistical} 
\begin{equation}
f(x;\alpha,\beta) = \frac{x^{\alpha-1} (1-x)^{\beta-1}}{\int_0^1 u^{\alpha-1} (1-u)^{\beta-1}},
\end{equation}
which are limited to $x \in [0,1]$ and normalized. By optimizing parameters $\alpha$ and $\beta$ using EM, we are able to find the likelihood of a measured population $h$ being assigned to a zero value $P(0|h)$. We classify a population $h$ as a zero if $P(0|h) > 0.5$ and as a peak otherwise. The results of this classification method for three different implemented function oracles with $r=1,2$ and $4$ can be found in Fig.~\ref{QPF_MM}. The classified non-zero values agree with the theoretical predictions, allowing us to correctly determine the period $r$ for all three implemented oracles.

\newpage

\end{document}